%% file: FrascatiLectures.tex
\documentclass[vecphys]{svmult}

\usepackage{makeidx}         
\usepackage{multicol}        
\usepackage[bottom]{footmisc}

\makeindex             

\begin{document}

\title*{The Attractor Mechanism in Five Dimensions}
\author{Finn Larsen}
\institute{Michigan Center for Theoretical Physics, Department of Physics \\ 
University of Michigan, Ann Arbor, MI-48109, USA. \\
\texttt{larsenf@umich.edu}}
%
%

\maketitle

\vspace*{1.5cm}

\centerline{\bf Abstract}
We give a pedagogical introduction to the attractor mechanism. We begin by 
developing the formalism for the simplest example of spherically symmetric black 
holes in five dimensions which preserve supersymmetry. We then discuss the 
refinements needed when spherical symmetry is relaxed. This is motivated by
rotating black holes and, especially, black rings. 
An introduction to non-BPS attractors is included, as is a discussion of thermodynamic 
interpretations of the attractor mechanism.\footnote{Lectures presented at the Winter School on the Attractor Mechanism (Frascati, March 20-24, 2006).}

\section{Introduction}
\label{sec:0}

These lectures are intended as a pedagogical introduction to the attractor
mechanism. With this mission in mind we will seek to be explicit and,
to the extent possible, introduce the various ingredients using rather
elementary concepts. While this will come at some loss in mathematical 
sophistication, it should be helpful to students who are not already 
familiar with the attractor mechanism and, for the experts, it may serve
to increase transparency. 

A simple and instructive setting for studying the attractor mechanism is M-theory 
compactified to five dimensions on a Calabi-Yau three-fold. The resulting low energy 
theory has $N=2$ supersymmetry and it is based on real special geometry. 
We will focus on this setting because of the pedagogical mission of the lectures: real 
special geometry is a bit simpler than complex special geometry, 
underlying $N=2$ theories in four dimensions. 

The simplest example where the attractor mechanism applies is that of a regular, 
spherically symmetric black hole that preserves supersymmetry. In the first lecture 
we develop the attractor mechanism in this context, and then verify the results by 
considering the explicit black hole geometry. 

In the second lecture we generalize the attractor mechanism to situations 
that preserve supersymmetry, but not necessarily spherical symmetry. 
Some representative examples are rotating black holes, multi-center black holes, 
black strings, and black rings. Each of these examples introduce new features 
that have qualitative significance for the implementation of the attractor mechanism. 
The approach will follow the paper \cite{Kraus:2005gh} rather closely, with the 
difference that here we include many more examples and other pedagogical material 
that should be helpful when learning the subject. 

In the third lecture we consider an alternative approach to the attractor 
mechanism which amount to seeing the attractor behavior as a result of an 
extremization procedure, rather than a supersymmetric flow. One setting that
motivates this view is applications to black holes that are extremal, but not 
supersymmetric. Extremization principles makes it clear that the attractor mechanism 
applies to such black holes as well. 

Another reason for the interest in extremization principles is more philosophical: 
we would like to understand what the attractor mechanism means in terms of 
physical principles. There does not yet seem to be a satisfactory formulation 
that encompasses all the different examples, but there are many interesting
hints. 

The literature on the attractor mechanism is by now enormous. As general references
let us mention from the outset the original works \cite{Ferrara:1995ih} establishing the 
attractor mechanism. It is also worth highlighting the review \cite{Moore:2004fg} which
considers the subject using more mathematical sophistication than we do here. 
In view of the extensive literature on the subject we will not be comprehensive 
when referencing. Instead we generally provide just a few references that may serve 
as entry points to the literature.

\section{The Basics of the Attractor Mechanism}
\label{sec:1}
In this section we first introduce a few concepts from the geometry of 
Calabi-Yau spaces and real special geometry. 
We then review the compactification of eleven-dimensional 
supergravity on a Calabi-Yau space and the resulting $N=2$ supergravity 
Lagrangian in five dimensions. This sets up a discussion of the attractor mechanism 
for spherically symmetric black holes in five dimensions. We conclude the lecture 
by giving explicit formulae in the case of toroidal compactification. 

\subsection{Geometrical Preliminaries}
On a complex manifold with hermitian metric $g_{\mu{\bar\nu}}$  
it is useful to introduce the K\"{a}hler two-form $J$ through
\begin{equation}
J = i g_{\mu{\bar\nu}} dz^\mu \wedge dz^{\bar\nu}~.
\end{equation}
K\"{a}hler manifolds are complex manifolds with hermitian metric 
such that the corresponding K\"{a}hler form is closed, $dJ=0$. 
The linear space spanned by all closed $(1,1)$ forms (modulo
exact forms) is an important structure that is known as the
Dolbault cohomology and given the symbol $H^{1,1}_{\bar\partial}$.
If we denote by $J_I$ a basis of this cohomology we can expand the 
closed K\"{a}hler form as 
\begin{equation}
\label{aexp}
J = X^I J_I ~~~~~;~I=1,\ldots, h_{11}~.
\end{equation}
This expansion is a statement in the sense of cohomology so it should be 
understood modulo exact forms.

Introducing the basis $(1,1)$-cycles $\Omega^I$ we can write the 
expression
\begin{equation}
\label{XIdef}
X^I = \int_{\Omega^I} J~~~~~; ~I=1,\ldots, h_{11},
\end{equation}
for the real expansion coefficients $X^I$ in (\ref{aexp}). We see that they
can be interpreted geometrically as the volumes of $(1,1)$-cycles within the 
manifold. The $X^I$ are known as K\"{a}hler moduli. In the context of 
compactification the K\"{a}hler moduli become functions on spacetime and 
so the $X^I$ will be interpreted as scalar fields. 

One of several ways to define a Calabi-Yau space is that it is a K\"{a}hler manifold 
that permits a globally defined holomorphic three-form. One consequence of this 
property is that Calabi-Yau spaces do not have any $(0,2)$ and $(2,0)$ forms. For this 
reason the $(1,1)$ cycles $\Omega^I$ are in fact the only two-cycles on the manifold. 

The two-cycles $\Omega^I$ give rise to a dual basis of four-cycles $\Omega_I$, 
$I=1,\ldots, h_{11}$, constructed such that their intersection numbers 
with the two-cycles are canonical $(\Omega^I,\Omega_J)=\delta^I_J$. The volumes 
of the four-cycles are measured by the K\"{a}hler form as
\begin{equation}
X_I = {1\over 2} \int_{\Omega_I} J\wedge J~.
\end{equation}
The integral can be evaluated by noting that the two-form $J_I$ covers 
the space transverse to the $4$-cycle $\Omega_I$. Therefore
\begin{equation}
\label{fourc}
X_I = {1\over 2} \int_{CY} J\wedge J\wedge J_I = {1\over 2} C_{IJK}
X^J X^K~,
\end{equation}
where the integrals 
\begin{equation}
C_{IJK} = \int_{CY} J_I \wedge J_J \wedge J_K~,
\end{equation}
are known as intersection numbers because they count the points
where the four-cycles $\Omega_I$, $\Omega_J$ and $\Omega_K$
all intersect. 

\subsection{The Effective Theory in Five Dimensions}
\label{effthy}
We next review the compactification of M-theory on a Calabi-Yau manifold \cite{Cadavid:1995bk}.
The resulting theory in five dimensions can be approximated at large distances 
by $N=2$ supergravity. In addition to the $N=2$ supergravity multiplet, the 
low energy theory will include matter organized into a number of $N=2$ 
vector multiplets and hypermultiplets. In discussions of the attractor mechanism 
the hyper-multiplets decouple and can be neglected. We therefore
focus on the gravity multiplet and the vector multiplets. 

The $N=2$ supergravity multiplet in five dimensions contains the metric, a vector 
field, and a gravitino (a total
of 8+8 physical bosons+fermions). Each $N=2$ vector multiplet in five dimensions
contains a vector field, a scalar field, and a gaugino (a total of 4+4 physical bosons+fermions). 
It is useful to focus on the vector fields. These fields all have their origin
in the three-form in eleven dimensions which can be expanded as
\begin{equation}
\label{gaud}
{\cal A} = A^I \wedge J_I~~~~;~I=1,\ldots, h_{11}~.
\end{equation}
The $J_I$ are the elements of the basis of $(1,1)$ forms introduced
in (\ref{aexp}). Among the
$h_{11}$ gauge fields $A^I$, $I=1,\ldots, h_{11}$, the linear combination
\begin{equation}
\label{graviphoton}
A^{\rm grav} = X_I A^I~,
\end{equation}
is a component of the gravity multiplet. This linear combination is
known as the graviphoton. The remaining $n_V = h_{11} -1$ vector fields are 
components of $N=2$ vector multiplets. 

The scalar components of the vector multiplets are essentially the scalar fields 
$X^I$ introduced in (\ref{XIdef}). The only complication is that, since one of
the vector fields does not belong to a vector multiplet, it must be that
one of the scalars $X^I$ also does not belong to a vector multiplet. 
Indeed, it turns out that the overall volume of the Calabi-Yau space
\begin{equation}
\label{constraint}
{\cal V} = {1\over 3!} \int_{CY} J\wedge J\wedge J = {1\over 3!} C_{IJK} X^I X^J X^K~,
\end{equation}
is in a hyper-multiplet. As we have already mentioned, hyper-multiplets decouple and
we do not need to keep track of them. Therefore (\ref{constraint}) can be treated as a 
constraint that sets a particular combination of the $X^I$'s to a constant. 
The truly independent scalars obtained by solving the constraint (\ref{constraint}) are 
denoted $\phi^i$, $i=1,\ldots, n_V$. These are the scalars that belong to vector multiplets.

We now have all the ingredients needed to present the Lagrangean of the theory.
The starting point is the bosonic part of eleven-dimensional supergravity 
\begin{equation}
S_{11} = {1\over 2\kappa_{11}^2} \int \left[ - R ~{}^*1 - {1\over 2}{\cal F} \wedge {}^* {\cal F} - {1\over 3!}
{\cal F} \wedge {\cal F} \wedge {\cal A} \right]~,
\end{equation}
where the four-form field strength is ${\cal F}=d{\cal A}$. The coupling constant
is related to Newton's constant as $\kappa^2_D = 8\pi G_D$. 
Reducing to five dimensions we find
\begin{equation}
\label{fivedact}
S_5
 = {1\over 2\kappa_5^2} \int 
 \left[ - R ~{}^*1 - G_{IJ} dX^I \wedge {}^* dX^J
 - G_{IJ} F^I \wedge {}^* F^J
 - {1\over 3!} C_{IJK} F^I \wedge F^J\wedge A^K\right]~,
 \end{equation}
where $F^I = dA^I$ and $\kappa_{5}^2=\kappa^2_{11}/{\cal V}$. The hodge-star is now the
five-dimensional one, although we have not introduced new notation to stress this fact.
 
The gauge kinetic term in (\ref{fivedact}) is governed by the metric
\begin{equation}
\label{GIJdef}
G_{IJ} = {1\over 2} \int_{CY} J_I \wedge {}^*J_J ~. 
\end{equation}
It can be shown that 
\begin{equation}
\label{bbaa}
G_{IJ} = - {1\over 2} \partial_I \partial_J (\ln {\cal V}) = -{1\over 2{\cal V}} (C_{IJK} X^K - 
{1\over{\cal V}}X_I X_J)~,
\end{equation}
where the notation $\partial_I = {\partial\over\partial X^I}$. 
Combining (\ref{fourc}) and (\ref{constraint}) we have the relation 
\begin{equation}
\label{bba}
X_I X^I=3{\cal V}~,
\label{xixi}
\end{equation} 
and so (\ref{bbaa}) gives
\begin{equation}
G_{IJ}X^J = {1\over 2{\cal V}} X_I~.
\label{gijlower}
\end{equation}
The metric $G_{IJ}$ (and its inverse $G^{IJ}$) thus lowers (and raises) the
indices $I,J=1,\cdots, h_{11}$. It is sometimes useful to extend this action
to the intersection numbers $C_{IJK}$ so that, {\it e.g.}, the constraint
(\ref{constraint}) can be be reorganized as
\begin{equation}
{\cal V}^2 = {1\over 3!}C^{IJK} X_I X_J X_K~,
\label{constrainttwo}
\end{equation}
where all indices were either raised or lowered.

The effective action in five dimensions
(\ref{fivedact}) was written in terms of the fields $X^I$ which include
some redundancy because the constraint (\ref{constraint}) should be imposed on them. 
An alternative form of the scalar term which employs only the unconstrained scalars
$\phi^i$ is
\begin{equation}
{\cal L}_{\rm scalar}  = -{1\over 2\kappa_5^2} g_{ij} d\phi^i \wedge {}^* d\phi^j~,
\end{equation}
where the metric on moduli space is
\begin{equation}
g_{ij} 
= G_{IJ} \partial_i X^I \partial_j X^J~.
\label{gijdef}
\end{equation}
Here derivatives with respect to the unconstrained fields are
\begin{equation}
\partial_i X^I= {\partial X^I\over\partial\phi^i} ~. 
\end{equation}

So far we have just discussed the bosonic part of the supergravity action. We will not need 
the explicit form of the terms that contain fermions. However, it is important that the 
full Lagrangean is invariant under the supersymmetry variations 
\begin{eqnarray}
\label{susyone}
\delta \psi_\mu &=&  \left[  D_\mu (\omega) + {i\over 24}
X_I (\Gamma_\mu^{~\nu\rho} - 4\delta^\nu_\mu \Gamma^\rho)F_{\nu\rho}^I \right] \epsilon~, \\
\label{susytwo}
\delta\lambda_i &=& -{1\over 2} G_{IJ} \partial_i X^I \left[ {1\over 2}
\Gamma^{\mu\nu} F_{\mu\nu}^J + i \Gamma^\mu \partial_\mu X^J \right] \epsilon~,
\end{eqnarray}
of the gravitino $\psi_\mu$ and the
gauginos $\lambda_i$, $i=1,\ldots, n_V$.
Here $\epsilon$ denotes the infinitesimal supersymmetry parameter. $D_\mu(\omega)$ is
the covariant derivative formed from the connection $\omega$ and acting on the spinor $\epsilon$. 
The usual $\Gamma$-matrices in five dimensions are denoted
$\Gamma^\mu$; and their multi-index versions $\Gamma^{\mu\nu}$ and $\Gamma^{\mu\nu\rho}$ 
are fully anti-symmetrized products of those. 

\subsection{A First Look at the Attractor Mechanism}
\label{attractor}
We have now introduced the ingredients we need for a first look at the attractor mechanism.
For now we will consider the case of supersymmetric black holes. As the terminology
indicates, such black holes preserve at least some of the supersymmetries. This means
$\delta \psi_\mu = \delta\lambda_i=0$ for some components of the supersymmetry
parameter $\epsilon$. A great deal can be learnt from these conditions by analyzing the 
explicit formulae (\ref{susyone}-\ref{susytwo}). 

In order to make the conditions more explicit we will make some simplifying assumptions.
First of all, we will consider only stationary solutions in these lectures. 
This means we assume that
the configuration allows a time-like Killing vector. The corresponding coordinate will be
denoted $t$. All the fields are independent of this coordinate. The
supersymmetry parameter $\epsilon$ satisfies
\begin{equation}
\label{susyproj}
\Gamma^{\hat t} \epsilon = -i\epsilon~,
\end{equation}
where hatted coordinates refer to a local orthonormal basis. In order to keep
the discussion as simple and transparent as possible we will for now also 
assume radial symmetry. This last assumption is very strong and will be relaxed 
in the following lecture. At any rate, under these assumptions 
the gaugino variation (\ref{susytwo}) reads
\begin{equation}
\label{az}
\delta\lambda_i = {i\over 2}G_{IJ} \partial_i X^I ( F^J_{m{\hat t}} - \partial_m X^J ) 
\Gamma^m \epsilon =0~,
\end{equation}
where $m$ is the spatial index. We exploited that, due to radial symmetry, only 
the electric components $F^I_{m{\hat t}}$ of the field strength can be nonvanishing. 
We next assume that the solution preserve $N=1$ supersymmetry so that (\ref{susyproj}) 
are the only projections imposed on the spinor $\epsilon$. Then $\Gamma^m\epsilon$
will be nonvanishing for all $m$ and the solutions to (\ref{az}) must satisfy
\begin{equation}
\label{aza}
G_{IJ} \partial_i X^I (F_{m{\hat t}}^J - \partial_m X^J ) =0~.
\end{equation}
This is a linear equation that depends on the bosonic fields alone. It
essentially states that the gradient of the scalar field (of type $J$) 
is identified with the electric field (of the same type). This identification is at
the core of the attractor mechanism. Later we will take more carefully
into account the presence of the overall projection operator 
$G_{IJ} \partial_i X^I$ in (\ref{aza}). This operator takes into account that fact that 
no scalar field is a superpartner of the graviphoton. This restriction arises here because 
the index $i=1,\ldots, n_V$ enumerating the gauginos is one short of the vector 
field index $I=1,\ldots, n_V+1$.

The conditions (\ref{aza}) give rise to an important monotonicity property that 
controls the attractor flow. To see this, multiply  by $\partial_r\phi^i$ and sum over $i$. 
After reorganization we find
\begin{equation}
\label{monot}
G_{IJ} \partial_r X^I F_{r{\hat t}}^J  = G_{IJ} \partial_r X^I  \partial_r X^J \geq 0~.
\end{equation}
The quantity on the right hand side of the equation is manifestly positively definite. 
In order to simplify the left hand of the equation we need to analyze Gauss' law for
the flux. For spherically symmetric configurations the Chern-Simons terms in the
action (\ref{fivedact}) do not contribute so the Maxwell equation is just
\begin{equation}
d(G_{IJ} {}^* F^J ) =0~.
\end{equation}
Using the explicit form of the metric for a radially symmetric
extremal black hole in five dimensions
\begin{equation}
\label{lineelm}
ds^2 = - f^{2} dt^2 + f^{-1} (dr^2 + r^2 d\Omega_3^2)~,
\end{equation}
the component form of the corresponding Gauss' law reads
\begin{equation}
\partial_r (G_{IJ} r^3 f^{-1} F^J_{r{\hat t}}) =0~.
\end{equation}
This can be integrated to give the explicit solution
\begin{equation}
\label{gauol}
G_{IJ} F^J_{r{\hat t}} = f\cdot{1\over r^3}\cdot {\rm const} \equiv f \cdot {Q_I \over r^3}~,
\end{equation}
for the radial dependence of the electric field. Inserting this in (\ref{monot}) we find the 
flow equation
\begin{equation}
\label{floweq}
\partial_r ( X^I Q_I ) = f^{-1} r^3 G_{IJ} \partial_r X^I \partial_r X^J \geq 0~.
\end{equation}
We can summarize this important result as the statement that the central charge
\begin{equation}
\label{zedef}
Z_e \equiv X^I Q_I~, 
\end{equation}
depends monotonically on the radial coordinate $r$. It starts as a
maximum in the asymptotically flat space and decreases as the black 
hole is approached. This is the attractor flow. 

In order to analyze the behavior of (\ref{floweq}) close to the horizon it
is useful to write it as
\begin{equation}
\label{floweqn}
r\partial_r Z_e = f^{-2} r^4 ~\epsilon  \geq 0~,
\end{equation}
where the energy density in the scalar field is 
\begin{equation}
\epsilon = g^{rr} G_{IJ} \partial_r X^I \partial_r X^J~.
\label{enerden}
\end{equation}
According to the line element (\ref{lineelm}) an event horizon at $r=0$ is characterized
by the asymptotic behavior $f\sim r^2$. Therefore the measure factor 
$f^{-2} r^4$ is finite there. Importantly, when $f\sim r^2$ the proper distance  to 
the horizon diverges as $\int_0 dr/r$. Since the horizon area is finite this means the proper 
volume of the near horizon region diverges. This is a key property of extremal black holes.
In the present discussion the important consequence is that the energy density of the scalars 
in the near horizon region must vanish, or else they would have infinite energy, and so 
deform the geometry uncontrollably. We conclude that the right hand side of (\ref{floweqn}) 
vanishes at the horizon, {\it i.e.} the inequality is saturated there. 
We therefore find the extremization condition
\begin{equation}
r\partial_r Z_e=0~,~~({\rm at~horizon})~.
\end{equation} 
This is the spacetime form of the attractor formula. 

There is another form of the attractor formula that is cast entirely in terms of the
moduli space. To derive it, we begin again from (\ref{aza}), simplify using Gauss' 
law (\ref{gauol}), and introduce the central charge (\ref{zedef}). We can write
the result as
\begin{equation}
\partial_i Z_e = \sqrt{g_\perp} g_{ij} \partial_n \phi^j~,
\label{modext}
\end{equation}
where $\sqrt{g_\perp}= f^{-3/2} r^3$ is the area element,
$g_{ij}$ is the metric on moduli space introduced in (\ref{gijdef}), 
$\partial_n = \sqrt{g^{rr}}\partial_r$ is the proper normal derivative, and the $\phi^j$ are the 
unconstrained moduli. As discussed in the previous paragraph, the energy density 
(\ref{enerden}) must vanish at the horizon for extremal black holes. 
This means the contribution from each of the unconstrained moduli must vanish by itself, 
and so the right hand side of (\ref{modext}) must vanish for all values of the index $i$. 
We can therefore write the attractor formula as an extremization principle over
moduli space
\begin{equation}
\partial_i Z_e=0~,~~({\rm at~horizon})~.
\label{modattr}
\end{equation} 
This form of the attractor formula determines the values $X^I_{\rm ext}$ of the scalar 
fields at the horizon in terms of the charges $Q_I$. 

We can solve ({\ref{modattr}) explicitly. In order to take the constraint (\ref{constraint}) 
on the scalars properly into account it is useful to rewrite the extremization principle
as
\begin{equation}
D_I Z_e=0~,~~({\rm at~horizon})~,
\label{modatt}
\end{equation} 
where the covariant derivative is defined as
\begin{equation}
D_I Z_e = \left(\partial_I-{1\over 3}(\partial_I\ln {\cal V})\right)Z_e
= (\partial_I - {1\over 3{\cal V}}X_I)Z_e = Q_I - {1\over 3{\cal V}}X_I Z_e~.
\label{covdef}
\end{equation}
We see that $Q_I\propto X_I$ at the attractor point, with the constant of proportionality determined 
by the constraint (\ref{constrainttwo}) on the scalar. We thus find the explicit result 
\begin{equation}
{X_{I}^{\rm ext}\over {\cal V}^{2/3}} = {Q_I \over \left( {1\over 3!} C^{JKL} Q_J Q_K Q_L\right)^{1/3} }  ~,
\label{ellattra}
\end{equation}
for the attractor values of the scalar fields in terms of the charges. 
As a side product we found 
\begin{equation}
{Z_e^{\rm ext}\over {\cal V}^{1/3}}  = 3 \left( {1\over 3!}C^{JKL} Q_J Q_K Q_L\right)^{1/3}~,
\label{expze}
\end{equation}
for the central charge at the extremum.

\subsection{A Closer Look at the Attractor Mechanism}
\label{moredetails}
Before considering examples, we follow up on some of the important features
of the attractor mechanism that we skipped in the preceding subsection: we 
introduce the black hole entropy, we discuss the interpretation of
the central charge, and we present some details on the units.

\subsubsection{Black Hole Entropy}

Having determined the scalars $X^I$ in terms of the charges we can now
express the central charge (\ref{zedef}) in terms of charges alone. It turns
out that for spherically symmetric black holes the resulting expression is
in fact related to the entropy through the simple formula
\begin{equation}
\label{bhent}
S = 2\pi\cdot{\pi\over 4G_5}\cdot\left( {1\over 3{\cal V}^{1/3}} Z_e^{\rm ext}\right)^{3/2}~.
\end{equation}
The simplest way to establish this relation is to inspect a few explicit black hole
solutions and then take advantage of near horizon symmetries to extend the
result to large orbits of black holes that are known only implicitly. The
significance of the formula (\ref{bhent}) is that it allows the determination of 
the black hole entropy without actually constructing the black hole geometry. 

In view of the explicit expression (\ref{expze}) for the central charge at the extremum
we find the explicit formula
\begin{equation}
\label{bhentexp}
S = 2\pi\cdot{\pi\over 4G_5}\cdot \sqrt{ {1\over 3!}C^{JKL} Q_J Q_K Q_L }~,
\end{equation}
for the black hole entropy of a spherically symmetric, supersymmetric black hole in five dimensions.

\subsubsection{Interpretation of the Central Charge}
In the preceding subsection
we introduced the central charge (\ref{zedef}) rather formally, as the
linear combination of charges that satisfies a monotonic flow. This characterization can be 
supplemented with a nice physical interpretation as follows. The eleven-dimensional
origin of the gauge potential $A^I_t$ can be determined from the decomposition (\ref{gaud}).
It is a three-form with one index in the temporal direction and the other two within the
Calabi-Yau, directed along a $(1,1)$-cycle of type $I$. Such a three-form is sourced by 
$M2$-branes wrapped on the corresponding $(1,1)$-cycle which we have denoted 
$\Omega^I$. The volume of this cycle is precisely $X^I$, according to (\ref{XIdef}). Putting
these facts together it is seen that the central charge (\ref{zedef}) is the total volume of the 
wrapped cycles, with multiple wrappings encoded in the charge $Q_I$. We can interpret
the underlying microscopics as a single $M2$-brane wrapping some complicated cycle
$\Omega$ within the Calabi-Yau which can be characterized in terms of a decomposition 
\begin{equation}
\Omega = Q_I \Omega^I~,
\end{equation}
on the canonical cycles $\Omega^I$. Then the central charge is identified with the mass
of this $M2$-brane, up to an overall factor of the tension. 

There is yet another interpretation of the central charge which takes as
starting point the $N=2$ supersymmetry algebra
\begin{equation}
\label{susyalge}
\{ Q^A_\alpha , Q^B_\beta \} = 2 \left(  \delta^{AB} P_\mu (\Gamma^\mu )_{\alpha\beta}
+ \delta_{\alpha\beta} \epsilon^{AB} Z_e \right)~,
\end{equation}
where $A,B=1,2$ distinguish the two supercharges. 
The last term on the right hand side (proportional to $Z_e$) is the 
central term. It is introduced from a purely algebraic point of view, 
as a term that commutes with all other generators of the algebra. 
The algebra is most usefully analyzed in the restframe where 
$P_\mu (\Gamma^\mu )_{\alpha\beta}=P_0 (\Gamma^0)_{\alpha\beta}$.
Consider a state that is annihilated by one or more of the supercharges $Q^A_\alpha$.
Taking expectation value on both sides with respect to this states,
and demanding positive norm of the state, we find
the famous BPS inequality
\begin{equation}
M = |P_0 | \geq Z_e~,
\end{equation}
with the inequality saturated exactly when supersymmetry is preserved by the state. 
Supersymmetric black holes are BPS states and so their mass should
agree with the algebraic central charge. In the preceding paragraph 
we showed that the mass agrees with the central charge introduced geometrically, 
so the alternate introductions of the central charge agree.

\subsubsection{Some Comments on Units and Normalizations}

Let us conclude this subsection with a few comments on units. It is standard to 
introduce the eleven dimensional Planck length through 
$\kappa_{11}^2 = (2\pi)^7 l_P^9$. In this notation the five dimensional Newton's constant 
is 
\begin{equation}
G_5 = {\pi\over 4}\cdot {(2\pi l_P)^6\over {\cal V}}\cdot l^3_P~,
\end{equation}
and the
$M2$-brane tension is $\tau_{M2} = {1\over (2\pi)^2 \ell_P^3}$.
The relation to standard string theory units are $l_P = g^{1/3}_s\sqrt{\alpha^\prime}$
and the radius of the M-theory circle is $R_{11}=g_s\sqrt{\alpha^\prime}$. Now,
the physical charges $Q_I$ were introduced in (\ref{gauol}) as the constant of 
integration from Gauss' law, following standard practice in supergravity. Such
physical charges are proportional to quantized charges $n_I$ according to
\begin{equation}
\label{chargeunit}
Q_I = ( {{\cal V}\over(2\pi l_P)^6})^{-2/3} \cdot l^2_P \cdot 
n_I = \left( {\pi\over 4G_5}\right)^{-2/3} n_I~.
\end{equation}
The mass of the brane configuration is
\begin{equation}
\label{massunit}
M = \tau_{M2} X^I n_I = 
{1\over l_P^3} \cdot {{\cal V}\over (2\pi l_P)^6} \cdot {X^I\over {\cal V}^{1/3}}\cdot Q_I
= {\pi\over 4G_5}\cdot {X^I\over {\cal V}^{1/3}}\cdot Q_I~.
\end{equation}
The formulae (\ref{chargeunit}-\ref{massunit}) are the precise versions of the informal
notions that the charge $Q_I$ counts the number of branes and that the central 
charge $Z_e$ agrees with the mass. We see that there are awkward constants
of proportionality, which vanish in units where $G_5 = {\pi\over 4}$ and the volumes of 
two-cycles are measured relative to ${\cal V}^{1/3}$. In this first lecture we will 
for the most part go through
the trouble of keeping all units around, to make sure that it is clear where the various
factors go. In later lectures we will revert to the simplified units.\footnote{
In fact, the supersymmetry algebra (\ref{susyalge}) was 
already simplified this way, to avoid overly heavy notation.} 
If needed, one can restore units by referring back to the simpler special cases. 

\subsection{An Explicit Example}
\label{example}
We conclude this introductory lecture by working out a simple example explicitly. 
The example we consider is when the Calabi-Yau space is just a torus $CY=T^6$. 
Strictly speaking a torus is not actually a Calabi-Yau space if by the latter we mean a
space with exactly $SU(3)$ holonomy. The issue is that for M-theory on $T^6$
the effective five-dimensional theory has $N=8$ supersymmetry rather than  
$N=2$ supersymmetry as we have assumed. This means there are extra 
gravitino multiplets in the theory which we have not taken into account. However,  
these gravitino multiplets decouple from the black hole background and so it is consistent 
to ignore them, in much the same way that we already ignore the $N=2$ hypermultiplets. 
We can therefore use the formalism reviewed above without any change. 

In the explicit example we will 
further assume that the metric on the torus is diagonal so that the K\"{a}hler form
takes the product form
\begin{equation}
\label{tormet}
J = i \left( 
X^1 dz^1\wedge d{\bar z}^1 + X^2 dz^2 \wedge d{\bar z}^2+ X^1 dz^3\wedge d{\bar z}^3 \right)~.
\end{equation}
Then the scalar fields $X^I$ with $I=1,2,3$ are just the volumes of 
each $T^2$ in the decomposition $T^6 = (T^2)^3$. The only nonvanishing intersection numbers 
of these two-cycles are $C_{123}=1$ (and cyclic permutations). 
The constraint (\ref{constraint}) on the scalars therefore takes the simple form
\begin{equation}
\label{torusc}
X^1 X^2 X^3 ={\cal V}~.
\end{equation}
The volumes (\ref{fourc})
of four-cycles on the torus are 
\begin{equation}
X_1 = X^2 X^3 = {\cal V}/X^1~~~({\rm and~cyclic~permutations})~.
\label{invvol}
\end{equation} 

\subsubsection{Attractor Behavior}

The central charge (\ref{zedef}) for this example is
\begin{equation}
\label{av}
Z_e = X^1 Q_1 + X^2 Q_2 + X^3 Q_3~.
\end{equation}
According to the extremization principle we can determine the scalar fields at the horizon
by minimizing this expression over moduli space. The constraint (\ref{torusc}) 
can be implemented by solving in terms of one of the $X^I$'s and then 
extremizing (\ref{av}) over the two remaining moduli. Alternatively, one can employ 
Lagrange multipliers. Either way the result for the scalars in terms of the charges is
\begin{equation}
\label{attrpt}
{X_1^{\rm ext}\over {\cal V}^{1/3}} = \left({Q_1^2\over Q_2 Q_3}\right)^{1/3}
= { Q_1 \over \left(Q_1 Q_2 Q_3\right)^{1/3} }~~~~~({\rm and ~cyclic ~permutations})~.
\end{equation}
These are the horizon values for the scalars predicted by the attractor mechanism. They
agree with the general formula (\ref{ellattra}). Below
we confirm these values in the explicit solutions. 

At the attractor point (\ref{attrpt}) the three terms in the central charge (\ref{av}) are 
identical. The central charge takes the value
\begin{equation}
Z_{\rm ext} = 3(Q_1 Q_2 Q_3)^{1/3}~.
\end{equation}
The black hole entropy (\ref{bhent}) becomes
\begin{equation}
\label{bhex}
S = 2\pi\cdot {\pi\over 4G_5}\cdot (Q_1 Q_2 Q_3 )^{1/2} = 
2\pi (n_1 n_2 n_3 )^{1/2}~.
\end{equation}
This is the entropy computed using the attractor formalism, {\it i.e.} without 
explicit construction of the black hole geometry. At the risk of seeming heavy
handed, we wrote (\ref{bhex}) both in terms of the proper (dimensionful) charges
$Q_I$ and also in terms of the quantized charges $n_I$. 

The entropy formula (\ref{bhex}) is rather famous so let us comment a little
more on the relation to other work. The $M2$-brane black hole considered here
can be identified, after duality to type IIB theory, with the $D1-D5$ black hole
considered by Strominger and Vafa \cite{Strominger:1996sh}. 
In this duality frame two of the M2-brane charges become the background D-branes 
and the third charge is the momentum $p$ along the $D1$-brane. Then (\ref{bhex}) 
coincides with Cardy's formula
\begin{equation}
S = 2\pi \sqrt{ch\over 6}~,
\end{equation}
where the central charge $c=6N_1 N_5$ for the CFT on the D-branes and $h=p$ for
the energy of the excitations. In the present lectures we are primarily interested in 
macroscopic features of black holes and no further details on the microscopic 
theory will be needed. For more review on this 
consult {\it e.g.} \cite{David:2002wn}.

\subsubsection{Explicit Construction of the Black Holes}

We can compare the results
from the attractor computation with an explicit construction of the black hole. 
The standard form of the $M2$-brane solution in eleven-dimensional supergravity
is
\begin{equation}
\label{M2sol}
ds^2_{11} = H^{-2/3} dx^2_\parallel
+ H^{1/3} dx^2_\perp~.
\end{equation}
Here the space parallel to the $M2$-brane is
\begin{equation}
dx^2_\parallel= -dt^2 + dx^2_1 + dx^2_2~,
\end{equation}
when the spatial directions of the $M2$-brane have coordinates $x_1$ and $x_2$.
The transverse space $dx^2_\perp$
is written similarly in terms of the remaining eight coordinates. The function $H$ can be 
any harmonic on the transverse space; the specific one needed in our example is given 
below.

The harmonic function rule states that composite solutions can be formed by
superimposing three $M2$-brane solutions of the form (\ref{M2sol}) with cyclically 
permuted choices of parallel space.
The only caveat is that we must smear along all directions within the torus, {\it i.e.} the 
harmonic functions can depend only on the directions transverse to all the different 
branes. This procedure gives the standard intersecting $M2$-brane solution
\begin{equation}
\label{intsol}
ds^2_{11} =  - f^2 dt^2 + f^{-1} (dr^2 + r^2 d\Omega^2_3)
+ \left[ \left({H_2 H_3 \over H^2_1}\right)^{1/3} (dx^2_1 + dx^2_2) + {\rm cyclic} \right]~,
\end{equation}
where
\begin{equation}
\label{fdef}
f = (H_1 H_2 H_3 )^{-1/3}~.
\end{equation}
We introduced radial coordinates in the four spatial dimensions transverse to all the branes. 
The harmonic functions are
\begin{equation}
\label{harmfct}
H_I = X_{I\infty} + {Q_I\over r^2}~~~~;~~I=1,2,3~.
\end{equation}
Comparing the intersecting brane solution (\ref{intsol}) with the torus metric (\ref{tormet})
we determine the scalar fields as
\begin{equation}
\label{XIex}
{X^1\over {\cal V}^{1/3}} = \left({H_2 H_3 \over H^2_1}\right)^{1/3}  ~~~~~({\rm and ~cyclic ~permutations})~.
\end{equation}
The only remaining matter fields from the five-dimensional point of view are the gauge 
fields 
\begin{equation}
\label{AIdef}
A^I = \partial_r H^{-1}_I dt~~~~~;~I=1,2,3~.
\end{equation} 

The scalar fields $X^I$ (\ref{XIex}) depend in a non-trivial way on the radial coordinate 
$r$. One can verify that the dependence is such that $Z_e=X^I Q_I$ is a monotonic 
function of the radii, but we will focus on the limiting values. 
The constants $X_{I\infty}$ in the harmonic functions (\ref{harmfct}) were introduced
in order to obtain the correct limit as $r\to\infty$
\begin{equation}
X^1 \to \left({X_{2\infty} X_{3\infty}\over   (X_{1\infty})^2}\right)^{1/3}  {\cal V}^{1/3} 
= X^1_\infty~~~~~({\rm and ~cyclic ~permutations})~.
\end{equation}
We used the constraint (\ref{torusc}) in the asymptotic space and the relation (\ref{invvol})
for the volumes of four-cycles. 
As the horizon ($r=0$)  is approached the moduli simplify to 
\begin{equation}
{X^1\over {\cal V}^{1/3}} \to { X^1_{\rm hor} \over {\cal V}^{1/3}}=  
\left({Q_2 Q_3\over Q_1^2}\right)^{1/3}  ~~~~~
({\rm and ~cyclic ~permutations})~.
\end{equation}
In view of (\ref{invvol}) this agrees with the values (\ref{attrpt}) predicted by the attractor 
mechanism.

We can also compute the black hole entropy directly from the geometry
(\ref{intsol}). The horizon at $r=0$
corresponds to a three-sphere with finite radius $R= (Q_1 Q_2 Q_3)^{1/6}$.
Since $V_{S^3} = 2\pi^2$ for a unit three-sphere
this gives the black hole entropy
\begin{equation}
S = {A\over 4G_5} = {1\over 4G_5} \cdot 2\pi^2\cdot R^3  = 2\pi (n_1 n_2 n_3 )^{1/2}~.
\end{equation}
This explicit result for the black hole entropy is in agreement with (\ref{bhex}) computed 
from the attractor mechanism.

\section{Black Ring Attractors}
\label{sec:2}
In this lecture we generalize the discussion of the attractor mechanism to a 
much larger class of stationary supersymmetric black solutions to the
$N=2$ theory in five dimensions introduced in section (\ref{effthy}). By giving up spherical
symmetry and allowing for dipole charges we can discuss multi-center black holes, 
rotating black holes and, especially, black rings.

\subsection{General Supersymmetric Solutions}
The most general supersymmetric metric with a time-like Killing vector is
\begin{equation}
\label{metric}
ds^2 = -f^2 (dt+\omega)^2 + f^{-1} ds^2_4 ~,
\end{equation}
where
\begin{equation}
\label{basemet}
ds^2_4 = h_{mn} dx^m dx^n~,
\end{equation}
is the metric of a four-dimensional base space and $\omega$ is a one-form on that
base space. In the simplest examples the base is just flat space, but generally it
can be any hyper-K\"{a}hler manifold in four dimensions. The matter fields needed
to support the solution are the field strengths $F^I=dA^I$ given by
\begin{equation}
\label{FIgen}
F^I = d( fX^I (dt+\omega)) + \Theta^I~,
\end{equation}
and the scalar fields $X^I$ satisfying the sourced harmonic equation
\begin{equation}
\label{XIgen}
{}^{(4)}\nabla^2(f^{-1} X_I) = {1\over 4} C_{IJK} \Theta^J \cdot \Theta^K~,
\end{equation}
on the base space. In these equations $\Theta^I$ is a closed self-dual two-form 
$\Theta^I ={}^{*_4}\Theta^I $ on  the base. This two-form vanishes in the
most familiar solutions but in general it must be turned on. For example,
it plays a central role for black rings. The inner product between
two-forms is defined as the contraction 
\begin{equation}
\label{contdef}
\alpha\cdot\beta = {1\over 2}\alpha_{mn} \beta^{mn}~.
\end{equation}
The self-dual part of the one-form $\omega$ introduced in the metric (\ref{metric}) 
is sourced by $\Theta^I$ according to
\begin{equation}
\label{omegadef}
d\omega  + {}^{*_4} d\omega = -f^{-1} X_I \Theta^I~.
\end{equation}

The general solution specified by equations (\ref{metric}-\ref{omegadef}) is a bit
impenetrable at first sight but things will become clearer as we study these
equations. At this point we just remark that the form of the solution given above
has reduced the full set of Einstein's equation and matter equations to a 
series of equations that are linear, if solved in the right order: first specify 
the hyper-K\"{a}hler base (\ref{basemet}) and choose a self-dual two 
form $\Theta^I$ on that base. Then solve (\ref{XIgen}) for $f^{-1}X_I$. 
Determine the conformal factor $f$ of the metric from the constraint (\ref{constraint})
and compute $\omega$ by solving 
(\ref{omegadef}). Finally the field strength is given in (\ref{FIgen}) \footnote{We 
need $X^I$ which can be determined from (\ref{fourc}). On a general Calabi-Yau
this is a nonlinear equation, albeit an algebraic one.}

\subsection{The Attractor Mechanism Revisited}
\label{attrmechrev}
We next want to generalize the discussion of the attractor mechanism from the
spherical case considered in section (\ref{attractor}) to the more general solutions
described above. Thus we consider the gaugino variations
\begin{eqnarray}
\delta\lambda_i &=& {i\over 2} G_{IJ} \partial_i X^I \left[ 
{i\over 2} F^J_{\mu\nu} \Gamma^{\mu\nu} - \partial_\mu X^J \Gamma^\mu \right] \epsilon~, \\
&=&  {i\over 2} G_{IJ} \partial_i X^I \left[ 
F^J_{m{\hat t}} \Gamma^m +{i\over 2} F^J_{mn}\Gamma^{mn}
-\partial_m X^J \Gamma^m \right]\epsilon~.
\label{magsusy}
\end{eqnarray}
In the second equation we imposed the supersymmetry projection (\ref{susyproj})
on the spinor $\epsilon$. In contrast to the spherically symmetric case 
(\ref{az}) there are in general both electric $E^I_m \equiv  F^I_{m{\hat t}}$ 
and magnetic $B^I_{mn}  \equiv  F^I_{mn}$ components of the field strength.
However, as we explain below, it turns out that the magnetic field in fact 
does not contribute to (\ref{magsusy}). Therefore we have
\begin{equation}
{i\over 2} G_{IJ} \partial_i X^I \left[ 
E^J_{m}-\partial_m X^J \right]\Gamma^m \epsilon =0~.
\end{equation}
Since this is valid for all components of $\epsilon$ we find 
\begin{equation}
\label{susyprr}
 G_{IJ} \partial_i X^I \left[ 
E^J_{m} 
-\partial_m X^J \right] =0~,
\end{equation}
just like (\ref{aza}) for the spherical symmetric case. In particular, we see that the
gradient of the scalar field is related to the electric field quite generally.
Of course this can be seen already from the explicit form (\ref{FIgen})
of the field strength, which can be written in components as
\begin{eqnarray}
E^I_m & \equiv&  F^I_{m{\hat t}} = f^{-1} \partial_m (fX^I) ~,
\label{elFI} \\
B^I_{mn} & \equiv&  F^I_{mn} = fX^I (d\omega)_{mn} + \Theta^I_{mn}~.
\label{magnFI}
\end{eqnarray}
The point here is that we see how the relation (\ref{elFI}) between the electric field
and the gradient of scalars captures an important part of the attractor mechanism
even when spherical symmetry is given up. 

The key ingredient in reaching this result was the claim that the magnetic
part (\ref{magnFI}) does not contribute to  the supersymmetry variation (\ref{magsusy}). 
It is worth explaining in more detail how this comes about. The first term in
(\ref{magnFI}) is of the form $F^I_{mn} \propto X^I (d\omega)_{mn}$. This term cancels 
from (\ref{magsusy}) because 
\begin{equation}
G_{IJ} \partial_i X^I X^J =0~,
\end{equation}
due to special geometry. Let us prove this.
Lowering the index using the metric (\ref{gijlower}) we can use (\ref{fourc}) to find
\begin{equation}
X_I \partial_i X^I = {1\over 2} C_{IJK} X^J X^K \partial_i X^I
= {1\over 3!} \partial_i ( C_{IJK} X^I X^J X^K ) =0~.
\end{equation}
due to the constraint (\ref{constraint}) on the scalars $X^I$. This is
what we wanted to show.

We still need to consider the second term in (\ref{magnFI}), the one taking the form
$F^I_{mn} \propto \Theta^I_{mn}$. This term cancels from the supersymmetry 
variation (\ref{magsusy}) because the supersymmetry projection (\ref{susyproj}) 
combines with self-duality of $\Theta^I_{mn}$ to give
\begin{equation}
\label{suprojj}
\Theta^I_{mn} \Gamma^{mn} \epsilon =0~.
\end{equation}
In order to verify this recall that the $SO(4,1)$ spinor representation can be
constructed from the more familiar $SO(3,1)$ spinor representation by 
including Lorentz generators from using the chiral matrix 
$\Gamma^4\equiv \gamma^5 = -i\gamma^0 \gamma^1 \gamma^2 \gamma^3$.
All spinors that survive the supersymmetry projection (\ref{susyproj}) 
therefore satisfy $\Gamma^{1234}\epsilon =\epsilon$ by construction and 
this means the $\Theta_{12}$ term in (\ref{suprojj}) cancels the $\Theta_{34}$ term, 
etc. 

After this somewhat  lengthy and technical aside we return to analyzing the 
conditions (\ref{susyprr}). Following the experience from the spherically
symmetric case we would like to trade the electric field for the charges, 
by using Gauss' law. The Lagrangean (\ref{fivedact}) gives the Maxwell equation
\begin{equation}
d( G_{IJ} {}^* F^J) = {1\over 2}C_{IJK} F^J \wedge F^K~,
\end{equation}
with the source on the right hand side arising from the Chern-Simons term. 
Considering the coefficient of the purely spatial four-form we find
Gauss'  law
\begin{equation}
\label{gauslaw}
\nabla^m (f^{-1} E_{mI} ) = -{1\over 8} C_{IJK} \Theta^J \cdot \Theta^K~.
\end{equation}
In arriving at this result  we must take into account off-diagonal terms in the
metric (\ref{metric}) due to the shift by $\omega$ of the usual time element $dt$.
These contributions cancel with the terms coming from the first term in the
field strength (\ref{magnFI}). Effectively this means only the term of
the form $F^I_{mn} \sim \Theta^I_{mn}$ remains and it is those terms that give rise
to the inhomogenous terms in (\ref{gauslaw}). The physical interpretation is that the 
electric field is sourced by a distributed magnetic field which we may
interpret as a delocalized charge density. 

We are now ready to derive the generalized flow equation. 
Multiplying  (\ref{susyprr}) by $\partial_n \phi^i$ and contract with the 
base metric $h^{mn}$ we find
\begin{equation}
\partial^m X^I E_{mI} = G_{IJ} \partial^m X^I \partial_m X^J~,
\end{equation}
which can be reorganized as 
\begin{equation}
\nabla^m(X^I f^{-1} E_{mI}) - X^I\nabla^m(f^{-1} E_{mI}) = f^{-1} G_{IJ}
\partial^m X^I \partial_m X^J~,
\end{equation}
and then Gauss' law (\ref{gauslaw}) gives
\begin{equation}
\nabla^m (X^I f^{-1} E_{mI}) = f^{-1} G_{IJ} \partial^m X^I \partial_m X^J
- {X^I\over 8} C_{IJK} \Theta^J \cdot \Theta^K~.
\label{floeqn}
\end{equation}
This is the generalized flow equation. In the case where $\Theta^I=0$ the right hand side is 
positive definite and then the flow equation generalizes the monotonicity 
property found in (\ref{floweq}) to many cases without radial symmetry. However, the most 
general case has nonvanishing $\Theta^I$ and such general flows are more complicated. 

\subsection{Charges}
In order to characterize the more general flows with precision, it is useful to be more 
precise about how charges are defined. 

Consider some bounded spatial region $V$. It is natural to define the electric charge 
in the region by integrating the electric flux through the boundary  $\partial V$ as 
\begin{equation}
\label{QIdef}
Q_I (V) = {1\over 2\pi^2} \int_{\partial V} dS f^{-1} n^m E_{mI}~,
\end{equation}
where $n^m$ is an outward pointing normal on the boundary. If we consider two 
nested regions $V_2\subset V_1$ we have 
\begin{equation} Q_I(V_1) - Q_I(V_2) = -{1\over 16\pi^2} \int d^4 x\sqrt{h}
C_{IJK} \Theta^J \cdot \Theta^K~,
\end{equation}
where the second step used Gauss' law (\ref{gauslaw}). This means the charge is 
monotonically decreasing as we move to larger volumes. The reason that it does not 
have to be constant is that in general the delocalized source on the right hand side 
of (\ref{gauslaw}) contributes.

The central charge is constructed from the electric charges by dressing them
with the scalar fields. It was originally introduced
in (\ref{zedef}) but, in analogy with the definition (\ref{QIdef}) of the electric
charge in a volume of space, we may dress the electric field by the scalars as well
and so introduce the central charge in a volume of space as
\begin{equation}
\label{Zdef}
Z_e(V) = {1\over 2\pi^2} \int_{\partial V} dS f^{-1} n^m X^I E_{mI}~.
\end{equation}
Considering again a nested set of regions we can use (\ref{floeqn}) 
to show that the central charge satisfies
\begin{equation}
Z_e(V_1) - Z_e(V_2) = {1\over 2\pi^2} \int d^4 x \sqrt{h}
\left[ f^{-1} G_{IJ} \nabla^m X^I \nabla_m X^J
- {1\over 8} C_{IJK} X^I \Theta^J \cdot \Theta^K \right]~.
\end{equation}
When $\Theta^I=0$ the central charge is monotonically increasing as we
move outwards. This generalizes the result from the spherically symmetric
case to all cases where the two-forms vanish. When the system is not
spherically symmetric there is no unique ``radius'' but this is circumvented by
the introduction of nested regions, which gives an orderly sense of
moving "outwards". Note that in general we do not force the nested
volumes to preserve topology. In particular there can be multiple singular 
points and these then provide natural centers of the successive nesting. 

When the two-forms $\Theta^I\neq 0$ the electric central charge (\ref{Zdef}) may 
not be monotonic and the flow equation does not provide any strong constraint
on the flow. 

In order to interpret the $\Theta^I$'s properly we would like to associate charges 
with them as well. Since they are two-forms it is natural to integrate them over 
two-spheres and so define  
\begin{equation}
\label{qdef}
q^I = -{1\over 2\pi} \int_{S^2} \Theta^I~.
\end{equation}
Since the two-forms $\Theta^I$ are closed the integral is independent under
deformations of the two-cycle and, in particular, it vanishes unless the $S^2$
is non-contractible on the base space. One way such non-trivial cycles can arise 
is by considering non-trivial base spaces. For our purposes the main example will
be when the base space is flat, but endowed with singularities along 
one or more closed curves (including lines going off to infinity). This situation also 
gives rise to noncontractible $S^2$'s because in four Euclidean dimensions a line can be 
wrapped by surfaces that are topologically a two-sphere. 

The charges $q^I$ defined in (\ref{qdef}) can be usefully thought as a magnetic 
charges. In our main example of a flat base space with a closed curve
we may interpret the configuration concretely in terms of electric charge distributed 
along the curve. Since the curve is closed there is in general no net electric charge, 
but there will be a dipole charge and it is this dipole charge that we identify as the 
magnetic charge (\ref{qdef}). 

In keeping with the analogy between the electric and magnetic charges we would 
also like to introduce a magnetic central charge. The electric central charge (\ref{Zdef}) 
was obtained by dressing the ordinary charge (\ref{QIdef}) by the moduli. 
In analogy, we construct the magnetic central charge
\begin{equation}
\label{Zmdef}
Z_m(V) = -{1\over 2\pi} \int_{S^2} X_I \Theta^I~.
\end{equation}
In some examples this magnetic central charge will play a role analogous to that played 
by the electric central charge in the attractor mechanism. 

A general configuration can be described in terms of its singularities on the 
base space. There may be a number of isolated point-like singularities, to which
we assign electric charges, and there may be a number of closed curves (including lines
going off to infinity), to which we assign magnetic charges. 

In four dimensions electric and magnetic charges are very similar: they are related
by electric magnetic duality, which is implemented by symplectic transformations in
the complex special geometry. In five dimensions the situation is more complicated because 
the Chern-Simons term makes the symmetry between point-like electric sources and
string-like magnetic sources more subtle. Therefore we will need to treat them independently. 

\subsection{Near Horizon Enhancement of Supersymmetry}
\label{susyenh}

There is another aspect of attractor behavior that we have not yet developed:
the attractor leads to enhancement of supersymmetry \cite{Chamseddine:1996pi}. 
This is a very strong condition that completely 
determines the attractor behavior, even when dipole charges are turned on. 

The enhancement of supersymmetry means the {\it entire} supersymmetry 
of the theory is preserved near the horizon. To appreciate why that is such a 
strong conditions, recall the origin of the attractor flow: we considered the 
gaugino variation (\ref{magsusy}) and found the flow by demanding that the 
various terms cancel. The enhancement of supersymmetry at the attractor 
means each term vanishes by itself. 

We first determine the supersymmetry constraint on the gravitino variation (\ref{susyone}).
By considering the commutator of two variations \cite{Chamseddine:1996pi}, it can be 
shown that the near horizon geometry must take the form $AdS_p\times S^q$. 
In five dimensions there are just two options: $AdS_3\times S^2$ or $AdS_2\times S^3$. 

The near horizon geometry of the supersymmetric black hole in five dimension that 
we considered in section (\ref{example}) is indeed $AdS_2\times S^3$ \cite{Strominger}
(up to global identifications). 
A more stringent test is the attractor behavior of the supersymmetric rotating black hole.
One might have expected that rotation would squeeze the sphere and make it oblate 
but this would not be consistent with enhancement of supersymmetry. In fact, it turns 
out that, for supersymmeric black holes, the near horizon geometry indeed remains 
$AdS_2\times S^3$ \cite{Cvetic:1999ja} (up to global identifications). 

There are also examples of a supersymmetric configurations with near horizon 
geometry $AdS_3\times S^2$. The simplest example is the black string in five 
dimensions. A more general solution is the supersymmetric black ring, 
which also has near horizon geometry $AdS_3\times S^2$. Indeed, the extrinsic
curvature of the ring becomes negligible in the very near horizon geometry so
there the black ring reduces to the black string. We will consider these 
examples in more detail in the next section. 

The pattern that emerges from these examples is that black holes correspond
to point-like singularities on the base and a near horizon geometry
$AdS_2\times S^3$ in the complete space. On the other hand, black strings 
and black rings correspond to singularities on a curve in the base and a near 
horizon geometry $AdS_3\times S^2$ in the complete space. The 
two classes of examples are related by electric-magnetic duality which, 
in five dimensions, interchanges one-form potentials with two-form potentials 
and so interchanges black holes and black strings. This duality 
interchanges $AdS_3\times S^2$ with $AdS_2\times S^3$.

So far we have just considered the constraints from the gravitino variation 
(\ref{susyone}). The attractor behavior of the scalars is controlled by the 
gaugino variation (\ref{magsusy}) which we repeat for ease of reference
\begin{equation}
\delta\lambda_i = 
{i\over 2} G_{IJ} \partial_i X^I \left[ 
F^J_{m{\hat t}} \Gamma^{m{\hat t}} +{i\over 2} F^J_{mn}\Gamma^{mn}
-\partial_m X^J \Gamma^m \right]\epsilon~.
\label{susythree}
\end{equation}
Near horizon enhancement of supersymmetry demands that each term in 
this equation vanishes by itself, since no cancellations are possible when
the spinor $\epsilon$ remains general. Let us consider the three conditions
in turn. 

The vanishing of the third term $\partial_m X^J=0$ means $X^J$ is a constant in 
the near horizon geometry. The attractor mechanism will determine the value of 
that constant as a function of the charges. 

The first term in (\ref{susythree}) reads 
\begin{equation}
\partial_i X^I E_{Im}=0~,
\end{equation}
in terms of the 
electric field introduced in (\ref{elFI}). In the event that there is a point-like singularity in 
the base space there is an $S^3$ in the near horizon geometry. Integrating the flux
over this $S^3$ and recalling the definition  (\ref{QIdef}) of the electric charge 
we then find
\begin{equation}
\partial_ i Z_e =0 ~,
\label{attzea}
\end{equation}
in terms of the electric central charge (\ref{zedef}). This is the attractor formula 
(\ref{modattr}), now applicable in the near any point-like singularity in base 
space. We can readily determine the explicit attractor behavior as (\ref{ellattra}) 
near any horizon with $S^3$ topology.

We did not yet consider the condition that the second term in (\ref{susythree}) 
vanishes. This term was considered in some detail after (\ref{magnFI}). There we found 
that the magnetic field $B^I_{mn}=F^I_{mn}$ has a term proportional to $X^I$ 
which cancels automatically from the supersymmetry conditions, due to special 
geometry relations. However, there is also another term in $B^I_{mn}$
which is proportional to $\Theta^I_{mn}$. This term also cancels from the supersymmetry
variation, but only for the components of the supersymmetry generator
$\epsilon$ that satisfy the projection (\ref{susyproj}). However, in the near horizon region 
there is enhancement of supersymmetry and so the variation must vanish for all 
components $\epsilon$. This can happen if the two-forms $\Theta^I$
take the special form \begin{equation}
\Theta^I= k X^I~,
\label{nearhTheta}
\end{equation}
where $k$ is a constant ($I$-independent) two-form because then
special geometry relations will again guarantee supersymmetry. The special 
form (\ref{nearhTheta}) will determine the scalars completely. 

Indeed, suppose that sources are distributed along a curve in the base space. Then 
we can integrate (\ref{nearhTheta}) along the $S^2$ wrapping the curve. This 
gives 
\begin{equation}
q^I = -{1\over 2\pi} \int_{S^2} \Theta^I = X^I_{\rm ext} \cdot {\rm constant}~,
\end{equation}
for the dipole charges in the near horizon region. The constant of proportionality is 
determined by the constraint (\ref{constraint}) and so we reach the final 
result\footnote{In this sections 3 and 4 we use the simplified units where ${\cal V}=1$
and $G_5={\pi\over 4}$. See section \ref{moredetails} for details on units.}
\begin{equation}
X^I_{\rm ext} = {q^I \over \left( {1\over 3!} C_{JKL} q^J q^K q^L\right)^{1/3}}~,
\label{magattr}
\end{equation}
for the scalar field in terms of the dipole charges. The result is applicable
near singularities distributed along a curve in the base space. In particular, 
this is the attractor value for the scalars in the near horizon region of black 
strings and of black rings. 

Our result (\ref{magattr}) was determined directly in the near horizon region, by exploiting 
the enhancement of supersymmetry there. In the case where $\Theta^I\neq0$
we cannot understand the entire flow as a gradient flow of the electric central charge
$Z_e$, nor are the attractor values given by extremizing $Z_e$. In fact, we can see
that the attractor values (\ref{magattr}) amount to extremization of the {\it magnetic}
central charge (\ref{Zmdef}). However, the significance of this is not so clear since it is
only the near horizon behavior that is controlled by $Z_m$, not the entire flow. It
would be interesting to find a more complete description of the entire flow in
the most general case. For now we understand the complete flow when $\Theta^I=0$,
and the attractor behavior when $\Theta^I\neq0$. 
 
There is in fact another caveat we have not mentioned so far. Our expression 
(\ref{ellattra}) for the scalars at the electric attractor breaks down when 
$C^{JKL} Q_J Q_K Q_L=0$, and similarly (\ref{magattr}) for the magnetic 
attractor breaks down when $C_{IJK} q^I q^J q^K = 0$. In the electric case
the issue has been much studied: the case where $C^{JKL} Q_J Q_K Q_L=0$ 
corresponds to black holes with area that vanishes classically. These are the small 
black holes. In some cases it is understood how higher derivative corrections to the 
action modify the attractor behavior such that the geometry and the attractor values 
of the scalars become regular \cite{LopesCardoso:1998wt,smallBHrefs}. 
The corresponding magnetic 
case $C_{IJK} q^I q^J q^K = 0$ corresponds to small black rings. This case has
been studied less but it is possible that a similar picture applies in that situation.

\subsection{Explicit Examples}
In this subsection we consider a number of explicit geometries. In each example we 
first determine the attractor behavior abstractly, by applying the attractor mechanism, 
and then check the results by inspecting the geometry. 

\subsubsection{The Rotating Supersymmetric Black Hole}
The simplest example of the attractor mechanism is the spherically symmetric black 
hole discussed in detail in section (\ref{example}). The generalization of the spherically 
symmetric solution to include angular momentum are the rotating supersymmetric black 
hole in five-dimensions. This solution is known as the BMPV black 
hole \cite{Breckenridge:1996is}. 

Let us consider the attractor mechanism first. The rotating black hole is electrically charged
but there are no magnetic charges, so the two-forms $\Theta^I$ vanish in this 
case. We showed in section (\ref{attrmechrev}) that then the electric central charge 
$Z_e$ must be monotonic just as it was in the nonrotating case. Extremizing over moduli 
space we therefore return to the values (\ref{attrpt}) of the scalars found in the 
non-rotating case. Alternatively we can go immediately to the general result (\ref{ellattra})
which is written for a general Calabi-Yau three-fold. Either way, we see that the 
attractor values of the scalars are independent of the black hole angular momentum. 
Since the rotation deforms the black hole geometry, this result is not at all obvious. 
The independence of angular momentum is a prediction of the attractor mechanism. 

We can verify the result by inspecting the explicit black hole solution. 
The metric takes the form (\ref{metric}) where the base space $dx^2_4$ is 
just flat space $R^4$ 
which can be written in spherical coordinates as
\begin{equation} 
dx^2_4 = dr^2 + r^2 (d\theta^2 + \cos^2\theta d\psi^2 + \sin^2\theta d\phi^2)~.
\end{equation}
Although the solution is rotating, it is almost identical to the non-rotating example 
discussed in section (\ref{example}): the conformal factor $f$ is given again by (\ref{fdef}) 
where the harmonic functions $H_I$ are given by (\ref{harmfct}). Additionally, the matter 
fields remain the scalar fields (\ref{XIex}) and the gauge fields (\ref{AIdef}). The only 
effect of adding rotation is that now the one-forms $\omega$ are 
\begin{equation}
\omega = -{J\over r^2}(\cos^2\theta d\phi + \sin^2\theta d\psi)~.
\end{equation}
As an aside we note that the self-dual part of $d\omega$ vanishes, as it must for solutions with
$\Theta^I=0$, but the anti-selfdual part is non-trivial: it carries the angular momentum. 

Now, for the purpose of the attractor mechanism we are especially interested in the scalar fields. 
As just mentioned, these take the form (\ref{XIex}) in terms of the harmonic
functions, independently of the angular momentum. This means they will in fact 
approach the attractor values (\ref{attrpt}) at the horizon. In particular,
the result is independent of the angular momentum, as predicted by the
attractor mechanism. 

\subsubsection{Multi-center black holes}
From the supergravity point of view, the $M2$-brane solution (\ref{M2sol})
is valid for {\it any} harmonic function $H$ on the transverse space. 
Similarly, the intersecting brane solution (\ref{intsol}) (and its generalization to an arbitrary
Calabi-Yau three-fold) remains valid for more general harmonic functions $H_I$. 
In particular, the standard harmonic functions (\ref{harmfct}) can be replaced by
\begin{equation}
H_I = X_{I\infty} + \sum_{i=1}^N {Q_I^{(i)}\over |\vec{r} - \vec{r}_i|^2}~.
\label{multiharm}
\end{equation}
where $\vec{r}_i$ are position vectors in the transverse space. We will assume
that all $Q_I^{(i)}>0$ so that the configuration is regular. 

The interpretation of these more general solutions is that they correspond
to multi-center black holes, {\it i.e.} $N$ black holes coexisting in equilibrium, 
with their gravitational attraction cancelled by repulsion of the charges. 
The black hole centered at $\vec{r}_i$ has charges $\{ Q_I^{(i)} \}$.

The attractor behavior of these solutions is the obvious generalization of the
single center black holes. The attractor close to each center depends only 
on the charges associated with that center, because the charge integrals (\ref{QIdef})
are defined with respect to singularities on the base manifolds. This immediately
implies that the attractor values for the scalars in a particular attractor region 
are (\ref{ellattra}) in terms of the charges $\{ Q_I^{(i)} \}$ associated with this
particular region. 

The explicit solutions verify this prediction of the attractor mechanism because 
the harmonic functions (\ref{multiharm}) are dominated by the term
corresponding to a single center in the attractor regime corresponding to that center. 

In some ways the multi-center solution is thus a rather trivial extension of
the single-center solution. The reason it is
nevertheless an interesting and important example is the following. Far from all 
the black holes, the geometry of the multi-center black hole approaches that 
of a single center solution with charges $\{ Q_I \} = \{ \sum_{i=1}^N Q_I^{(i)} \}$. Based
on the asymptotic data alone one might have expected an attractor flow governed 
by the corresponding central charge $Z_e = X^I Q_I$, leading to the attractor values 
for the scalars depending on the $Q_I$ in a unique fashion, independently of the 
partition of the geometry into constituent black holes with charges $\{ Q_I^{(i)} \}$. 
The multi-center black hole demonstrates that this expectation is false: the 
asymptotic behavior does {\it not} uniquely specify the attractor values of the scalars, 
and nor does it define the near horizon geometry and the entropy. 

More structure appears when one goes beyond the focus on attractor
behavior and consider the full attractor flow of the scalars. As we discussed in section 
(\ref{attrmechrev}), the flow of the scalars is a gradient flow 
controlled by the electric central charge (this is when the dipoles vanish). The
central charge is interpreted as the total constituent mass. For generic values
of the scalar fields the actual mass of the configuration is smaller, {\it i.e.} the black holes
are genuine bound states. Now, in the course of the attractor flow, the values
of the scalars change. At some intermediate point it may be that the actual
mass of the black hole is identical to that of two (or more) clusters of constituents. 
This is the point of marginal stability. There the attractor flow will split up, and continue
as several independent flows, each controlled by the appropriate sets of smaller
charges. This process then continues until the true attractor basins are reached.
The total flow is referred to as the split attractor flow. It has interesting
features which are beyond the scope of the present lecture. We refer the reader
to the original papers \cite{Denef:2000ar} and the review \cite{Moore:2004fg}.

\subsubsection{Supersymmetric Black Strings}
The black string is a five dimensional solution that takes the form
\begin{equation}
ds^2_5 = f^{-1} (-dt^2 +dx^2_4) + f^2(dr^2 + r^2 d\Omega^2_2)~,
\label{strmetric}
\end{equation}
where the conformal factor
\begin{equation}
f =  {1\over 3!} C_{IJK} H^I H^J H^K~,
\end{equation}
in terms of the harmonic function
\begin{equation}
H^I = X^I_{\infty} + {q^I\over 2r}~.
\end{equation}
The geometry is supported by the gauge fields
\begin{equation}
A^I =  - {1\over 2} q^I (1 + \cos\theta) d\phi~,
\label{strai}
\end{equation}
and the scalar fields
\begin{equation}
X^I = f^{-1} H^I~.
\label{strsoln}
\end{equation}

The black string solution is the long distance representation of an M5-brane that
wraps the four-cycle $q^I \Omega_I$ inside a Calabi-Yau threefold and has the 
remaining spatial direction aligned with the coordinate $x^4$. This configuration 
plays in important role in microscopic considerations of the four dimensional black hole
(see {\it e.g.} \cite{MSW}).
 
The gauge field (\ref{strai}) corresponds to the field strength 
$F^I = - q^I \sin\theta d\theta d\phi$. This is a magnetic field,
with normalization of the charge in agreement with the one 
introduced in (\ref{qdef}). The black string is therefore an example where
the two-forms $\Theta^I\neq 0$. 

We should note that the metric (\ref{strmetric}) of the supersymmetric black string 
differs from the form
(\ref{metric}), assumed in the analysis in this lecture. The reason
that a different form of the metric applies is that 
the black string has a null Killing vector whereas (\ref{metric}) assumes a 
time-like Killing vector. Nevertheless, we can think of the null case
as a limiting case of the time-like one. Concretely, if there
is a closed curve on the base-space of (\ref{metric}), the black string is the limit
where the curve is deformed such that two points are taken to infinity and
only a straight line remains ({\it i.e.} the return line is fully at infinity). This limiting
procedure is how the simple black string arises from the more complicated 
black ring solution (see following example). 

Let us now examine the attractor behavior of the black string. In section (\ref{susyenh}) 
we showed that near horizon enhancement of supersymmetry demands that, at the 
attractor, the two forms simplify to $\Theta^I= k X^I$ where $k$ is a 
constant ($I$ independent) two-form. This condition was then showed to imply the 
expression (\ref{magattr}) for the scalars as functions of the magnetic charges. 

We can verify the attractor behavior by inspection of the explicit solution. 
Taking the limit $r\to 0$ on the scalars (\ref{strsoln}) we find 
\begin{equation}
X^I_{\rm hor} = {q^I\over ({1\over 3!} C_{JKL} q^J q^K q^L)^{1/3}}~.
\end{equation}
This agrees with (\ref{magattr}) predicted by the attractor mechanism. 

\subsubsection{Black Rings}
As the final example we consider the attractor behavior near the supersymmetric
black ring \cite{Elvang:2004ds,Bena:2004de}. 
This is a much more involved example which in fact was the motivation for 
the development of the formalism considered in this lecture. 

The supersymmetric black ring is charged with respect to both electric charges
$Q_I$ and dipole charges $q^I$. Far from the ring the geometry is dominated
by the electric charges, which have the slowest asymptotic fall-off, and the value of the
charges can be determined using Gauss' law (\ref{QIdef}). The dipole charges
are determined according to (\ref{qdef}) where the by $S^2$ is wrapped around 
the ring. Since the two-forms do not vanish they dominate the near horizon geometry
and the near horizon values of the scalar fields become (\ref{magattr}), as 
they were for the black string.

We can verify the result from the attractor mechanism by inspecting the explicit 
black ring solution. The metric takes the general form (\ref{metric}). The conformal 
factor $f$ is given by (\ref{fdef}) in terms of functions $H_I$ which take the form:
\begin{equation}
H_I = X_{I\infty} + {Q_I - {1\over 2} C_{IJK} q^J q^K\over\Sigma}
+ {1\over 2} C_{IJK} q^J q^K {r^2\over\Sigma^2}~,
\end{equation}
where
\begin{equation}
\Sigma = \sqrt{ (r^2 - R^2)^2 + 4R^2 r^2 \cos^2\theta}~.
\end{equation}
Although $H_I$ play the same role as the harmonic functions in other examples they
are in fact not harmonic: they satisfy equations with sources. The expression for
$\Sigma$ vanishes when $r=R$, $\theta={\pi\over 2}$, and arbitrary $\psi$.
Therefore the functions $H_I$ diverge along a circle of radius $R$ in the base space. 
This is the ring. 

The full solution in five dimension remains regular, due to the conformal factor. 
At large distances $H_I\sim X_{I\infty} + {Q_I \over r^2}$ so the black ring 
has the same asymptotic behavior as the spherically symmetric black hole 
considered in section (\ref{example}). This is because the dipole charges
die off asymptotically and so $H_I$ differ from that of a black hole only at order
${\cal O}({1\over r^4})$. However, the dipole charges dominate close to the horizon. 

The scalar fields in the supersymmetric black ring solution take the form 
\begin{equation}
X_I = {H_I \over  ({1\over 3!}C^{JKL}H_J H_K H_L)^{1/3}}~.
\end{equation}
In the near horizon region where the "harmonic" functions $H_I$ diverge the
scalars approach
\begin{equation}
X^I = { q^I\over ({1\over 3!}C_{JKL} q^J q^K q^L )^{1/3}}~.
\end{equation}
This is in agreement with the prediction (\ref{magattr}) from the attractor mechanism. 

In the preceding we defined just enough of the black ring geometry to consider
the attractor mechanism. For completeness, let us discuss also the remaining 
features. They are most conveniently introduced in terms of the ring coordinates
\begin{equation}
h_{mn} dx^m dx^n = {R^2 \over (x-y)^2}
\left[ {dy^2 \over y^2 -1}
+ (y^2 -1) d\psi^2
+ {dx^2\over 1-x^2} + (1-x^2) d\phi^2 \right]~,
\end{equation}
on the base space. Roughly speaking, the $x$ coordinate is a polar 
angle $x\sim\cos\theta$ that combines with $\phi$ to form two-spheres in the
geometry. The angle along the ring is $\psi$, and $y$ can be interpreted as
a radial direction with $y\to -\infty$ at the horizon. In terms of these coordinates 
the two form sources are
\begin{equation}
\Theta^I = -{1\over 2} q^I ( dy\wedge d\psi + dx\wedge d\phi)~.
\end{equation}
Integrating the expression along the $S^2$'s we can verify that
the normalization agrees with the definition (\ref{qdef}) of magnetic charges. 

The final element of the geometry is the one-form $\omega$ introduced in (\ref{metric}).
Its nonvanishing components are
\begin{eqnarray}
\omega_\psi &=& -{1\over R^2} (1-x^2) \left[ 
Q_I q^I - {1\over 6} C_{IJK} q^I q^J q^K (3 + x + y)\right]~,
\label{omegaphi} \\
\omega_\phi &=& {1\over 2} X_{I\infty} q^I  (1+y) + \omega_\psi~.
\label{omegapsi}
\end{eqnarray}
In five dimensions there are two independent  angular momenta which we can 
choose as $J_\phi$ and $J_\psi$. The one form (\ref{omegaphi}-\ref{omegapsi}) 
gives their values as
 \begin{eqnarray}
J_\phi &=& {\pi\over 8G_5} ( Q_I q^I  - {1\over 6}C_{IJK} q^I q^J q^K )~,
\label{Jphi}  \\
J_\psi &=& {\pi\over 8G_5} ( 2R^2 X_{I\infty} q^I
+ Q_I q^I - {1\over 6}C_{IJK} q^I q^J q^K )~.
\label{Jpsi}
\end{eqnarray}
These expressions will play a role in the discussion of the interpretation of
the attractor mechanism in the next section.

\section{Extremization Principles}
\label{sec:3}
An alternative approach to the attractor mechanism is to analyze the Lagrangian
directly, without using supersymmetry \cite{GibKal}. An advantage of this method is that 
the results apply to all extremal black holes, not just the supersymmetric ones \cite{trivetal}. 
A related issue is the understanding of the attractor mechanism in terms of the 
extremization of various physical quantities.

\subsection{The Reduced Lagrangian}
The attractor mechanism can be analyzed without appealing to supersymetry, by starting 
directly from the Lagrangian. In this subsection we exhibit the details. 

We will consider just the spherically symmetric case with the metric
\begin{equation}
\label{sphmetric}
ds^2 = -f^2 dt^2 + f^{-1} (dr^2 + r^2 d\Omega^2_3)~.
\end{equation}
Having assumed spherical symmetry, it follows that the gauge field strengths take the form 
(\ref{gauol}). The next step is to insert the {\it ansatz} into the Lagrangian (\ref{fivedact}).
The result will be a reduced Lagrangian that depends only on the radial variable. In order 
to advantage of intuition from elementary mechanics, it is useful to trade the radial coordinate 
for an auxiliary time coordinate defined by
\begin{equation}
dr = - {1\over 2}r^3 d\tau~~~~;~~~\partial_r = -{2\over r^3}\partial_\tau~.
\end{equation}
Introducing the convenient notation
\begin{equation}
f=e^{2U}~,
\end{equation}
a bit of computation gives the reduced action
\begin{equation}
Ld\tau =   \left[ - 6(\partial_\tau U)^2 -  G_{IJ}\partial_\tau X^I \partial_\tau X^J
 + {1\over 4}e^{4U}G^{IJ}Q_I Q_J\right]d\tau~,
\label{redlag}
\end{equation}
up to overall constants. 

Imposing a specific {\it ansatz} on a dynamical system removes numerous degrees of 
freedom. The corresponding equations of motion appear as constraints on the 
reduced system. In the present setting the main issue is that the charges specified 
by the {\it ansatz} are the momenta conjugate to the gauge fields. The correct 
variational principle is then obtained by a Legendre transform which, in this simple case,
simply changes the sign of the potential in (\ref{redlag}). Thus the equations of motion
of the reduced system can be obtained in the usual way from the effective
Lagrangean
\begin{equation}
{\cal L} =   \left[ - 6(\partial_\tau U)^2 - G_{IJ}\partial_\tau X^I \partial_\tau X^J
- {1\over 4}e^{4U}G^{IJ}Q_I Q_J\right]~. 
\label{redham}
\end{equation}

It is instructive to rewrite the effective potential in (\ref{redlag}) and (\ref{redham}). 
Using the relations (\ref{xixi}-\ref{gijlower}) we can show the identity
\begin{equation}
G^{IJ}Q_I Q_J = {2\over 3} Z^2_e
+ G^{IJ} D_I Z_e D_J Z_e~,
\label{rewpot}
\end{equation}
where we used the definition  (\ref{zedef}) of the electric central charge $Z_e$ and
(\ref{covdef}) of the covariant derivative on moduli space. The Lagrangean (\ref{redham}) 
can be written as
\begin{eqnarray}
{\cal L} &=&    - 6(\partial_\tau U)^2 -  G_{IJ}\partial_\tau X^I \partial_\tau X^J
- {1\over 6}e^{4U}Z_e^2 
- {1\over 4} e^{4U}G^{IJ} D_I Z_e D_J Z_e\\
&=& -6 \left( \partial_\tau U \pm {1\over 6} e^{2U} Z_e \right)^2
\label{compsq} \\
&-& G_{IJ} \left(\partial_\tau X^I \pm {1\over 2} e^{2U} G^{IK} D_K Z_e \right)
\left(\partial_\tau X^J \pm {1\over 2} e^{2U} G^{JL} D_L Z_e \right)
\pm \partial_\tau \left(e^{2U} Z_e \right)~, \nonumber 
\end{eqnarray}
where we used\footnote{We can verify this by writing  
$D_I Z_e  = {\cal V}^{1/3} \partial_I ({\cal V}^{-1/3}Z_e )$.
This amounts to changing
into physical coordinates before taking the derivative and then changing back.} 
\begin{equation}
\partial_\tau X^I D_I Z_e  = \partial_\tau Z_e ~.
\end{equation}
Thus the Lagrangean can be written a sum of squares, up to a total derivative. We can 
therefore find extrema of the action by solving the linear equations of motion
\begin{eqnarray}
\partial_\tau U &=& -{1\over 6} e^{2U} Z_e~, \\
\partial_\tau X^I &=& -{1\over 2} e^{2U} G^{IJ} D_J Z_e~. 
\end{eqnarray}
The second equation is identical to the condition (\ref{modext}) that the gaugino variations 
vanish, as one can verify by identifying variables according 
to the various notations we have introduced. The first equation can be interpreted as 
the corresponding condition that the gravitino variation vanish. To summarize, we
have recovered the conditions for supersymmetry by explicitly writing the 
bosonic Larangean as a sum of squares, so that extrema can be found by solving certain 
linear equations of motion. The analysis of these linear equations can now be 
repeated from section \ref{attractor}. In particular, finite energy density at the horizon
(or enhancement of supersymmetry, as discussed in section \ref{susyenh}) 
implies the conditions $D_I Z_e=0$, and these in turn lead to the explicit form (\ref{expze})
for the attractor values of the scalars.

One of the advantages of this approach to the attractor mechanism is that it applies
even when supersymmetry is broken. To see this, consider solutions with constant value of 
the scalar fields throughout spacetime $\partial_\tau X^I=0$. Extremizing the Lagrangian
with respect to the scalar fields can then be found by considering just the potential (\ref{rewpot}).
Upon variation we find
\begin{equation}
\left( {2\over 3} G_{IJ} Z_e + D_I D_J Z_e \right) D^J Z_e=0~.
\end{equation}
This equation is solved automatically for $D_J Z_e=0$. Such geometries are
the supersymmetric solutions that have been our focus. However, it is seen that there 
can also be solutions where the scalars satisfy
\begin{equation}
{2\over 3} G_{IJ} Z_e + D_I D_J Z_e =0~.
\end{equation}
Such solutions do not preserve supersymmetry, but they do exhibit attractor 
behavior.

\subsection{Discussion: Physical Extremization Principles}
In section \ref{susyenh} we found that the attractor values are determined by
extremizing one of the two central charges. For $\Theta^I=0$ they
are determined by extremizing the electric central charge (\ref{zedef}) over 
moduli space $\partial_i Z_e=0$. On the other hand, for $\Theta^I\neq 0$,
we should instead extremize the magnetic central charge $\partial_i Z_m=0$.
These prescriptions are mathematically precise but they lack a clear physical 
interpretation. It would be nice to reformulate the extremization principles in terms 
of physical quantities.  

Let us consider first the situation when $\Theta^I=0$. As discussed in 
section \ref{moredetails}, the electric central charge can be interpreted as the 
mass of the system. Therefore, extremization amounts to minimizing the mass. 
If we think about the attractor mechanism this way, the monotonic flow of the 
electric central charge amounts to a roll down a potential, with scalars
ultimately taking the value corresponding to dynamical equilibrium.
In particular, if the scalars are adjusted to their attractor values already at infinity 
(these configurations are referred to as "double extreme black holes")
there is no flow because the configuration remains in its equilibirum. 

A difficulty with this picture is the fact that the situation with $\Theta^I\neq 0$ 
works very differently even though the asymptotic configuration is in fact 
independent of the dipole charges. We would like a physical extremization 
principle that works for that case as well. 
The case where $\Theta^I\neq 0$ is elucidated by 
considering the combination
\begin{equation}
J_\psi - J_\phi = R^2 X_{I\infty} q^I = R^2 Z_m~,
\label{angdiff}
\end{equation}
of the angular momenta (\ref{Jphi}-\ref{Jpsi}). This quantity can be interpreted 
as the intrinsic angular momentum of the black ring, not associated with the 
surrounding fields. The interesting point is that extremizing $Z_m$ is the same 
as extremizing $J_\psi - J_\phi$ with $R^2$ fixed. It may at  first seem worrying 
that we propose extremizing angular momenta. For a black hole these would 
be quantum numbers measurable at infinity, and so they would be part of the 
input that specifies solution. However, the black ring solution is different: we 
can choose its independent parameters as $q^I$, $Q_I$, $R^2$ with the 
understanding that then the angular momenta $J_\phi$ and $J_\psi$ that 
support the black ring must be those determined by (\ref{Jphi}-\ref{Jpsi}).
The precise values of $J_\phi$ and $J_\psi$ so determined depend on 
the scalars and the proposed extremization principle is that the scalars
at the horizon are such that the combination (\ref{angdiff}) is minimal. 

The proposed principle is quite similar to the extremization of the mass in the 
electric case of supersymmetric black holes. In fact, the combination (\ref{angdiff}) 
of  angular momenta that we propose extremizing in the magnetic case 
behaves very much like a mass: it can be interpreted as the momentum along 
the effective string that appears in the near ring limit \cite{ringent,myringent}. 

In order to elevate the extremization of (\ref{angdiff}) to a satisfying principle 
one would need a geometric definition of the ring radius $R$ that works 
independently of the explicit solution. Ideally, there should be some kind
of conserved integral, akin to those defining the electric charges,
or the more subtle ones appearing for dipole 
charges \cite{Copsey:2005se}. Another issue 
is that of more complicated multiple ring solutions, which are characterized by
several radii. This latter problem is completely analogous to the ambiguity with 
assigning mass for multi black hole solutions: the asymptotics does not uniquely 
specify the near horizon behavior. We will put these issues aside for now,
and seek an extremization principle that combines the extremization of (\ref{angdiff}) 
in the magnetic case with extremization of the mass in electric case, and 
works in any basin of attraction, whether electric or magnetic in character.

To find such a principle, recall that the black hole entropy (\ref{bhent}) can be 
written in terms of the central charge in the electric case. Accordingly, the 
extremization over moduli space can be recast as\footnote{Although
(\ref{bhent}) was given in the spherically symmetric case, it can be generalized
to include angular momentum \cite{Kallosh:1996vy} (just subtract $J^2$ under the
square root. The argument given below carries
through.}
\begin{equation}
\partial_i S =0~.
\label{entext}
\end{equation}
The black ring entropy can be written compactly as 
\begin{equation}
S = 2\pi \sqrt{J_4}~.
\end{equation}
For toroidal compactification\footnote{This statement has an obvious alternate 
version that applies to general Calabi-Yau spaces.} $J_4$ is the quartic $E_{7(7)}$ 
invariant, evaluated at arguments that depend on the black ring parameters according to the
identifications
\begin{equation} 
J_4 = J_4 (Q_I, q^I, J_\psi - J_\phi)~.
\end{equation}
The black ring is thus related to black holes in four dimensions \cite{ringent}.

In the present context the point is that the extremization principle
(\ref{entext}) applies to both electric and magnetic attractors. This
provides a thermodynamic interpretation of the attractor mechanism. 
One obstacle to a complete symmetry between the electric and
magnetic cases is that near a magnetic attractor point one must 
apply (\ref{entext}) with $Q_I$, $q^I$, and $R$ fixed, while near an 
electric attractor it is $Q_I$ and $J$ that should be kept fixed. In either case 
these are the parameters that define the solution. 

There is one surprising feature of the proposed physical extremization principle: 
the entropy is {\it minimized} at the attractor point. This may be the correct physics: 
as one moves closer to the horizon, the geometry is closer to the microscopic 
data. It is also in harmony with the result that, at least in some cases, extremization 
over the larger moduli space that includes multi-center configurations gives split 
attractor flows that correspond to independent regions that have even less 
entropy \cite{Denef:2000ar}, with the end of the flow plausibly corresponding 
to "atoms" that have no entropy at all \cite{Mathur:2005zp,atomization}. 

We end with a summary of this subsection: we have proposed an extremization 
principle (\ref{entext}) that applies to both the electric (black hole) and magnetic 
(black ring) cases. A physical interpretation in terms of thermodynamics looks
promising at the present stage of development. In order to 
fully establish the proposed principle one would need a more detailed
understanding of general flows, including those that have magnetic charges, 
and one would also need a more general definition of charges.

\acknowledgement
I thank Stefano Bellucci for organizing a stimulating meeting and 
Per Kraus for collaboration on the material presented in these
lectures. I also thank Alejandra Castro for reading the manuscript
carefully and proposing many improvements, and Josh Davis for
discussions.

%
%
%
\input{referenc}



\printindex
\end{document}

%% file: referenc.tex
%
%

%
%